\documentclass{ifacconf}

\usepackage{graphicx}      
\usepackage{natbib}        
\usepackage{epsfig, latexsym, amsmath, color, amsfonts, amssymb, diagbox,multirow,mathrsfs,amsfonts,dsfont}

\newtheorem{theorem}{Theorem}[section]
\newtheorem{lemma}[theorem]{Lemma}

\newtheorem{problem}[theorem]{Problem}
\newtheorem{definition}[theorem]{Definition}

\newcommand{\E}{{\mathbb{E}}}
\newcommand{\tr}{{\mathtt{Tr}}}
\newcommand{\pr}{{\mathtt{Pr}}}
\newcommand{\set}{{\{1,\cdots,N\}}}

\newcommand\addtag{\refstepcounter{equation}\tag{\theequation}}

\begin{document}
\begin{frontmatter}

\title{Multi-sensor Transmission Management\\ \mbox{for Remote State Estimation under Coordination}}

\thanks[footnoteinfo]{
}

\author[First]{Kemi Ding}
\author[Second]{Yuzhe Li}
\author[Third]{Subhrakanti Dey}
\author[First]{Ling Shi}

\address[First]{Department of Electronic and Computer Engineering,
Hong Kong University of Science and Technology, Hong Kong. (e-mail: \{kdingaa,eesling\}@ust.hk).}
\address[Second]{Department of Electrical and Computer Engineering, University of Alberta, Canada. (e-mail: pku.ltracy@gmail.com)}
\address[Third]{Department of Engineering Science,
Uppsala University, Sweden. (e-mail: subhrakanti.dey@angstrom.uu.se)}

\begin{abstract}                
This paper considers the remote state estimation in a cyber-physical system (CPS) using multiple sensors. The measurements of each sensor are transmitted to a remote estimator over a shared channel, where simultaneous transmissions from other sensors are regarded as interference signals.
In such a competitive environment, each sensor needs to choose its transmission power for sending data packets taking into account of other sensors' behavior. To model this interactive decision-making process among the sensors, we introduce a multi-player non-cooperative game framework.
To overcome the inefficiency arising from the Nash equilibrium (NE) solution, we propose a correlation policy, along with the notion of correlation equilibrium (CE). An analytical comparison of the game value between the NE and the CE is provided, with/without the power expenditure constraints for each sensor. Also, numerical simulations demonstrate the comparison results.

\end{abstract}


\end{frontmatter}

\section{Introduction}

Cyber-physical systems (CPSs), which combine the traditional control system with information and communication technologies, can provide great improvements in the system performance, including the robustness to unexpected disturbance and efficient utilization of resources, see \cite{6176187}. 
As the next generation control systems, CPSs have attracted increasing interest in different realms, such as the smart grid, intelligent transportation, ubiquitous health care, and so on.


The incorporation of communication networks although provides stability and efficiency for physical systems, unfortunately raises a number of technical challenges in the control system design.
For example, when we consider remote state estimation using multiple sensors, if the communication bandwidth is limited and cannot allow all sensors to transmit data, then simultaneous data transmission will lead to signal interference which will further lead to packet drop and hence deteriorate the estimation performance.
There are several representative methods for interference management in communication theory, such as code division multiple access (CDMA), see \cite{tse2005fundamentals}. However, there lack efficient approaches to cope with the multi-access issue in the remote state estimation.
Another factor to consider is the limited sensor energy budget.
As most sensor nodes use on-board batteries, which are difficult to replace or recharge, the energy for sensing, computation and transmission is restricted.
Motivated by this, a considerable amount of literature has been published on sensor transmission scheduling to achieve accurate estimation under limited energy constraints, e.g., \cite{shi2011time, ren2014dynamic}. However, many of them focus on the one-sensor case and model the sensor scheduling as a Markov decision problem (MDP). The problem becomes difficult when taking multiple sensors into account. In this work, we provide quantitative analysis of transmission competition over a shared channel for remote estimation under abundant energy and limited energy,respectively.


In communication theory, the traditional way to solve the competition problem is to model it as a non-cooperative game {(see \cite{CDMA2001Basar,koskie2005nash,machado2008survey,sengupta2010game})}. Precisely, the sensors are treated as selfish players aiming at {maximizing their utilities such as their own throughput or certain thresholds of signal-to-noise-ratio}, and the Nash equilibrium (NE) concept provides the optimal strategy for each player. Different from these preliminary works, our work focuses on dynamic systems and considers the state estimation performance. Since the sensors have different time-varying objective functions, more thorough analysis of the NE solution is required, as demonstrated in \cite{yuzhe2014multi}. Unfortunately, the obtained NE in \cite{yuzhe2014multi} leads to an inefficient outcome, called ``tragedy of the commons".
To overcome these limitations, we introduce the notion of correlated equilibrium (CE), along with a correlation mechanism, and analyze its impact on the state estimation performance.

As proposed by \cite{aumann1974subjectivity}, the CE is a generalization of the NE concept to capture the strategic correlation opportunities that the players face. More importantly, it allows an increase in all players' profits simultaneously. The definition of CE includes an arbitrator who can send (private or public) signals to the players. Remarkably, this arbitrator requires no intelligence or any knowledge of the system, which is different from centralized management (where everyone obeys some rules provided by the mediator). Therefore, the generated signal does not depend on the system states; for example, see \cite{han2012game}, the surrounding weather conditions or the thermal noises for communication channels. Unlike the cooperative game, each player is self-enforced to comply with the outcome suggested by the mediator, rather than being restricted by a contract. In conclusion, the CE concept not only provides a tractable solution for this competition problem, but also may bring more benefits to each player than the NE.


By the employment of the CE concept, we investigate the optimal transmission strategies for the sensors in a large-scale CPS with shared public resources.
Compared with the previous work by \cite{yuzhe2014multi}, not only is the performance difference between CE and NE studied, but the respective power constraint for each sensor is also considered. The main contributions of our current work are summarized as follows:
\begin{itemize}
  \item We provide a general game-theoretic framework for remote state estimation in a multi-access system, where the sensors compete over access to the same channel for packet transmission.
  \item In the absence of power restrictions, we analyze the existence and uniqueness of NE for this game. That is, at the NE, each sensor transmits with its maximum energy level. Moreover, the CE is proved to be equivalent to the NE.
  \item With energy limitations, we formulate the problem as a constrained game and provide the closed-form NE. Moreover, after proposing an easy-to-implement correlation mechanism, we obtain the explicit representation of the CE. By comparison, the CE is preferable to the NE for this game.
\end{itemize}

The remainder of the paper is organized as follows. Mathematical models of the system are described in Section \ref{section:problem setup}. In Section \ref{section:multi-sensor transmission game}, we introduce the multi-player non-cooperative game and give the definition of CE. Section \ref{section:main result} demonstrates the main theoretical results with/without power constraints. The correlation policy is also introduced in Section \ref{section:main result}, and the simulation results are shown in Section \ref{section:simulation and example}. Some concluding remarks are given in the end.

\textit{Notations}:
$\mathbb{Z}$ is the set of non-negative integers. $\mathbb{N}$ is the set of positive integers. $k\in \mathbb{Z}$ is the time index. $\mathbb{R}^{n}$ is the $n$ dimensional Euclidean space. $\mathbb{S}_{+}^{n}$ is the set of $n$ by $n$ positive semi-definite matrices. When $X \in \mathbb{S}_{+}^{n}$, it is written as $X \geq 0$. $X \geq Y$ if $X - Y \in  \mathbb{S}_{+}^{n}$. $\E[\cdot]$ is the expectation of a random variable and $\tr(\cdot)$ is the trace of a matrix. For functions $f, f_1, f_2: \mathbb{S}_{+}^{n} \rightarrow \mathbb{S}_{+}^{n}$, $f_1\circ f_2$ is defined as $f_1\circ f_2(X) \triangleq f_1\big(f_2(X)\big)$. $\mathds{1}(\cdot)$ is the indicator function. $\Delta(\cdot)$ represents a set of probability measures and ``w.p." means with probability.

\section{Problem Setup}\label{section:problem setup}

\begin{figure}
\begin{center}
\includegraphics[width=8.6cm]{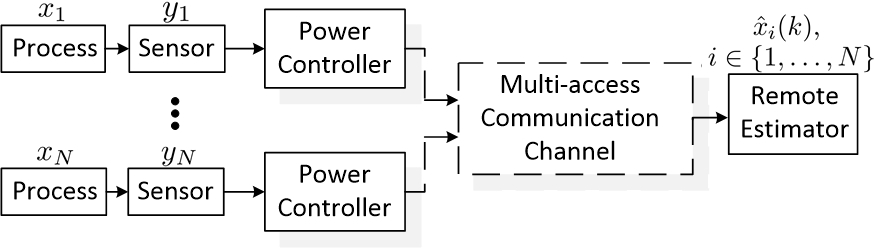}    \vspace{-3mm}
\caption{System Architecture.}
\label{fig:system}
\end{center}
\end{figure}

As depicted in Fig.~\ref{fig:system}, the state information of different processes is sent to the remote estimator through one shared channel, and essential components of the overall system structure will be introduced in this section.

\subsection{Local Kalman Filter}
Consider the following network system containing one remote estimator and $N$ sensors, which separately monitor different linear systems:
\begin{eqnarray}
x_{i}(k+1)  &=&  A_{i}x_i(k) + w_i(k), \label{eqn:process-dynamics} \\
y_{i}(k)  &=&  C_ix_i(k) + v_i(k), \; i\in\{1,\ldots,N\}, \label{eqn:measurement-equation}
\end{eqnarray}
where at time $k$, the state vector of the system measured by sensor $i$ is $x_i(k) \in \mathds{R}^{n_x}$, and the obtained noisy measurement is $y_i(k) \in \mathds{R}^{m_y}$. For each process $i\in \{1,\ldots,N\}$, the process noise $w_i(k) \in \mathds{R}^{n_x} $ and the measurement noise $v_i(k) \in \mathds{R}^{m_y}$ are zero-mean i.i.d. Gaussian random variables with $\mathbb{E}[w_i(k)w_{i}(j){'}] = \delta_{kj}Q_i$ ($Q_i\geq 0$), $\mathbb{E}[v_i(k)v_i(j){'}] = \delta_{kj}R_i$ ($R_i> 0$), and $\mathbb{E}[w_i(k)v_i(j){'}] = 0 \; \forall j,k$. The initial state $x_i(0)$ is a zero-mean Gaussian random vector with covariance $\Sigma_i(0)\geq 0$, and it is uncorrelated with $w_i(k)$ and $v_i(k)$. The time-invariant pair $(A_i, C_i)$ is assumed to be detectable and $(A_i,\sqrt{Q_i})$ is stabilizable.


Here, we adopt ``smart" sensors to improve the estimation/control performance of the current system. As illustrated in \cite{hovareshti2007sensor}, the so-called ``smart" sensors, equipped with memory and the embedded operators, are capable of processing the collected data. In Fig.~\ref{fig:system}, by running a Kalman filter locally, sensor $i$ can compute the optimal estimate of the corresponding state $x_i(k)$ based on the collected measurements $\{y_i(1),\ldots,y_i(k)\}$. The obtained minimum mean-squared error (MMSE) estimate of the process state is given by
$$\hat{x}^s_i(k)=\E[x_i(k)|y_i(1),\ldots,y_i(k)].$$
The corresponding estimation error covariance is denoted as:
$$P^s_i(k)  \triangleq  \E[(x_i(k) - \hat{x}^s_i(k)) (x_i(k) - \hat{x}^s_i(k))'|y_i(1),\ldots,y_i(k)].$$

These terms are computed recursively following the standard Kalman filter equations:
\begin{eqnarray*}
\hat{x}^s_i(k|k-1) & = & A_i\hat{x}^s_i(k-1), \\
P_i(k|k-1)^{s}  &=&  A_iP^s_i(k-1)A_i' + Q_i,  \\
K_i(k) & = & P^{s}_i(k|k-1)C_i'[C_iP^{s}_i(k|k-1)C_i' + R_i]^{-1},  \\
\hat{x}^s_i(k) & = & A_i\hat{x}^s_i(k-1) + K_i(k)y_i(k) - C_iA_i\hat{x}^s_i(k-1), \\
P^{s}_i(k) & = &(I -K_i(k)C_i)P^{s}_i(k|k-1).
\end{eqnarray*}
The iteration starts from $\hat{x}^s_i{0} = 0$ and $P^s_i(0) = \Sigma_i(0)$. For notational simplicity, we define the Lyapunov and Riccati operators $h_i(\cdot)$ and $\tilde{g}_i(\cdot): \mathds{S}_{+}^{n} \rightarrow \mathds{S}_{+}^{n}$ as
\begin{align*}
    h_i(X) &\triangleq A_iXA_i' + Q_i, \\
    \tilde{g}_i(X) &\triangleq X - XC_i'[C_iXC_i' + R_i]^{-1}C_iX.
\end{align*}

Suppose that the time-invariant pair $(A, C)$ is detectable and $(A,\sqrt{Q})$ is stabilizable, the estimation error covariance $P_k^s$ converges exponentially to a unique fixed point $\overline{P}_i$ of $h_i \circ \tilde{g}_i$ according to~\cite{anderson1979optimal}. For brevity, we ignore the transient periods and assume that the Kalman filter at the sensor has entered steady state; i.e.,
\begin{equation}\label{eqn:steady-state-assumption}
P^s_i(k) = \overline{P}_i,~k \geq 1.
\end{equation}

As mentioned in \cite{shi2011time}, the steady-state error covariance $\overline{P}_i$ has the following property.
\begin{lemma} \label{pro:error covariance}
For $0\leq t_1 < t_2$, the following inequality holds:
\begin{equation}
\tr[\overline{P}_i] \leq \tr[h^{t_1}_i(\overline{P}_i)] < \tr[ h^{t_2}_i(\overline{P}_i)].
\end{equation}
\end{lemma}

\subsection{Communication Model}
As demonstrated in Fig. \ref{fig:system}, the sensor $i$ will transmit the local estimate $\hat{x}^{s}_i(k)$ as a packet to the remote estimator through a single channel, which may be occupied by other sensors. Hence, this information delivery interferes directly with other transmissions of the sensors that use the same channel, hence, the estimation of state $x_i(k)$ is affected.

In this multi-access system, we assume that the shared channel has independent Additive White Gaussian Noise (AWGN). By modeling the signals of other sensors as interfering noises, the channel quality, as measured by the signal-to-noise-ratio (SNR) in point-to-point communication, is closely related to the revised signal-to-interference-and-noise-ratio (SINR), see~\cite{tse2005fundamentals}. For sensor $i\in\{1,\ldots,N\}$, its SINR is defined as:
\begin{equation}\label{eqn:SINR}
  \gamma_i(k) = L \frac{h_i a_i(k)}{\sum_{j\neq i}h_j a_j(k) + \sigma^2} = L \frac{q_i(k)}{\hat{q}_{-i}(k)+\sigma^2},
\end{equation}
in which $a_i(k)\geq 0$ and $a_j(k)\geq 0$ correspond to the transmission power taken by sensor $i$ and sensor $j$, respectively. For simplicity, we define $q_i(k)\triangleq h_i a_i(k)$ and $\hat{q}_{-i}(k) \triangleq \sum^{N}_{j=1,j\neq i} q_j(k)$. The extra term $\sum_{j\neq i}h_j a_j(k)$ in the denominator of (\ref{eqn:SINR}) is due to the interference from the other sensors, and $\sigma^2$ is the channel noise. The parameter $h_i \in (0,1), \forall i \in \{1,\ldots,N\}$ is the channel gain from sensor $i$ to the remote estimator, and $L$ is the spreading gain of the communication system. These channel parameters are assumed to be time-invariant\footnote{Time-variety has little effect on the results afterwards.}. Moreover, they can be acquired by the sensors, as the sensors can access the channel state information (CSI) using pilot-aided channel estimation techniques.

To characterize the packet-dropout for sensor $i$, we introduce the notation symbol error rate (SER) and adopt a general function to represent the relationship between SER and SINR:
$$ \text{SER}_i\triangleq f(\gamma_i), \; \forall i\in\set, $$
where $f(\cdot)$ depends on the channel characteristic and the modulation schemes. Note that since the interference takes great importance in the multi-access system, the packet transmission typically operates at low SINRs. As investigated by \cite{5437418}, the error probability function $f(\cdot)$ is strictly concave and decreasing in $\gamma_i$.

Here, we consider an erasure channel that the packet will be dropped if it contains any error (in general, the symbol error can be detected by the channel coding method). Therefore, the simultaneous transmissions of this system are characterized by independent Bernoulli processes, denoted by $\eta_i(k)$. Let $\eta_i(k)=0$ denotes the loss of packet $\hat{x}^{s}_i(k)$, and $\eta_i(k)=1$ otherwise. Hence, we have
$$ \pr(\eta_i(k)=1) = 1- f(\gamma_i), \;\forall i\in \set.$$
Note that, the arrival of packet $\hat{x}^{s}_i(k)$ not only depends on the transmission power of sensor $i$, but also is affected by the behaviors of the other sensors.

%
%

\subsection{Remote State Estimation}

Let $\hat{x}_i(k)$ denote the MMSE estimate of the process $x_i(k)$ generated by the remote estimator, with the error covariance matrix $P_i(k)$. Similar to \cite{shi2011time}, the estimation process is as follows: if
$\hat{x}^s_i(k)$ arrives successfully, the estimator synchronizes its respective estimate $\hat{x}_i(k)$ with it; otherwise, the estimator simply predicts the estimate based on its previous estimate and system dynamics. In short, the estimation $\hat{x}_i(k)$ is denoted by
$$\hat{x}_i(k) =  \eta_i(k) \hat{x}^s_i(k) + (1-\eta_i(k)) A_i\hat{x}_i(k-1).$$

Similarly, the simple recursion of the error covariance $P_i(k)$ is
\begin{eqnarray}\label{eqn:error for remote}
 P_i(k) & \triangleq & \E[(x_i(k)-\hat{x}_i(k)) (x_i(k)-\hat{x}_i(k))'] \nonumber \\
 &=& \eta_i(k) \overline{P}_i + (1-\eta_i(k)) h_i(P_i(k-1)),
\end{eqnarray}
where $\overline{P}_i$ stands for the steady-state error covariance defined in (\ref{eqn:steady-state-assumption}). For each sensor, we define a random variable $\tau_i(k) \in \mathds{Z}$ as the holding time:
\begin{equation}\label{eqn:holding time}
  \tau_i(k) \triangleq k - \max_{0\leq l \leq k}\{l: \eta_i(l) = 1 \},
\end{equation}
which represents the intervals between the present moment $k$ and the most recent time when the data packet $\hat{x}^{s}_i(k)$ arrives successfully. Without loss of generality, for all $i\in \set$, we assume that the initial packets $\hat{x}^s_i(0)$ are received by the estimator, and hence $\tau_i(0) = 0$.

Note that, the equivalent relationship between the holding time and the estimation error covariance at the remote estimator is
\begin{equation}\label{eqn:relationship between time and error}
P_i(k) = h_i^{\tau_i(k)}(\overline{P}_i).
\end{equation}
Furthermore, the iteration of the holding time is
\begin{equation}\label{eqn:holding time iteration}
 \tau_i(k+1) =  (1-\eta_i(k))( \tau_i(k)+1). \\
\end{equation}

\subsection{Problem of Interest}

In our work, every sensor competes for public communication resources to obtain an accurate estimation performance, which can be formulated as a game with multiple self-interested players. Alternatively, it can be modeled as a constrained game with the consideration of power limitations. The best response for each player is the NE in a traditional manner. Differently, in this work we consider the notion of CE and investigate whether the coordination (CE) mechanism, compared to the NE, can bring extra benefits to each player simultaneously.

\section{Multi-sensor Transmission Game}\label{section:multi-sensor transmission game}

In this section, we model the interactive decision-making process of each sensor as a multi-player game and introduce the concept of equilibrium solution.

\subsection{Game theoretic framework}

The multi-player game, denoted by $\mathcal{G}$, is characterized by a triplet $<\mathcal{I},\mathcal{A},\mathcal{U}>$ where

\subsubsection{Players:}

$\mathcal{I}= \set$ is the set of players, in which $i\in \mathcal{I}$ represents sensor $i$. As a necessary condition for the equilibrium analysis, we assume that all sensors are rational; that is, each sensor will make the best decision to maximize their benefit among all available actions. Also, the rationality assumption is common knowledge shared among the players.

\subsubsection{Actions:}

$\mathcal{A} = \{\mathcal{A}_i,i\in\mathcal{I}\}$ illustrates the set of actions for each player $i$. For simplicity, we consider the transmission action sets with discrete energy levels, i.e., $\mathcal{A}_i = \{e^{(1)}_i,\ldots, e^{(m_i)}_i\}$ for player $i$. Let $a_i(k) \in \mathcal{A}_i$ denote the transmission action (or pure strategy) taken by player $i$ at time $k$. The mixed strategy for each player, denoted by $s_i(k) \in \Delta(\mathcal{A}_i), i\in\set$, is a probability distribution over the pure action space $\mathcal{A}_i$. That is, player $i$, following strategy profile $s_i(k)$, may take the transmission power $e^{(l)}_i,l\in\{1,\ldots,m_i\}$ w.p. $s_{i}(a_i(k)=e^{(l)}_i)$\footnote{Here, we abuse the notation $s_{i}(a_i(k))$ to represent one element of vector $s_{i}(k)$.} at time $k$. Moreover, define $\textbf{a}(k) = \{a_1(k), a_2(k),\ldots,a_N(k)\}$ as the joint action played by the overall players. Alternatively, $\textbf{a}(k) = \{a_i(k),\textbf{a}_{-i}(k)\}$, in which $\textbf{a}_{-i}(k)$ represents the joint action excluding player $i$. Similarly, the joint strategy profiles are represented by $\textbf{s}(k) = \{s_1(k), s_2(k),\ldots, s_N(k)\} = \{s_i(k),\textbf{s}_{-i}(k)\}$.

%

\subsubsection{Utility:}

$\mathcal{U} = \{u_i,i\in\mathcal{I}\}$ is the utility set and $u_i$ represents the utility function for player $i$ with $u_i: \mathcal{A} \rightarrow \mathbb{R}$.
As discussed previously, each sensor focuses on improving its respective estimation accuracy, measured by the estimation error covariance. Hence, based on (\ref{eqn:error for remote}) the utility function for player $i$ is characterized by\footnote{In the rest of this paper, we will omit the variable $k$ of $\tau_i(k)$, $\gamma_i(k)$, $a_i(k)$, $s_i(k)$ and $q_i(k)$ when the underlying time index $k$ is obvious from the context; otherwise, it will be indicated.}
\begin{align*}\addtag\label{eqn:utility function}
u_i(\textbf{a}) &\triangleq -\tr\{\E[P_i(k)]\} \\
    &= -\tr\{\E[(1-f(\gamma_i))\overline{P}_i+f(\gamma_i)h_i^{\tau_i+1}(\overline{P}_i)]\} \\
    &= f(\gamma_i)c_i - \tr\{\E[\overline{P}_i]\},
\end{align*}
in which $c_i \triangleq \tr\{ \E[\overline{P}_i-h_i^{\tau_i+1}(\overline{P}_i)]\}$ is independent of $\gamma_i$, and $c_i<0$ is derived from Lemma \ref{pro:error covariance}.

Next, we define the expected utility function of player $i$ in a slight abuse of notation $u_i(\cdot)$. Under the joint strategy profile $\textbf{s}$, the benefit obtained by player $i$ is
\begin{align*}\addtag\label{eqn:expected utility}
  u_i(\textbf{s}) &\triangleq \sum_{\textbf{a}\in \mathcal{A}} \pr(\textbf{a}|\textbf{s}) u_i(\textbf{a}),
\end{align*}
in which $\pr(\textbf{a}|\textbf{s})$ is the probability over the joint action $\textbf{a}$ under strategy $\textbf{s}$.

\subsection{Equilibrium and Coordination}

In our current game, player $i$ is subject to maximize selfishly its utility at time $k$, i.e.,
\begin{problem}\label{prob:game problem}
For any player $i\in\mathcal{I}$,
\begin{align*}
\max_{\textbf{s}\in \Delta(\mathcal{A})} \; &u_{i}(\textbf{s}) \\
s.t. \; & \sum_{a_i \in \mathcal{A}_i} s_{i}(a_i) =1.
\end{align*}
\end{problem}

Note that the game theory, see \cite{fudenberg1991game}, provides a way to cope with these coupled optimization problems. One common solution concept is defined as follows:
\begin{definition}[Nash Equilibrium]\label{def:NE}
In this multi-player one-stage game with finite action space, the strategy profile $\textbf{s}^{\star,NE} =\{s_1^{\star,NE},\ldots, s_N^{\star,NE}\}$ is a Nash equilibrium if no player can benefit from changing strategies while the others keep their own equilibrium strategy unchanged; i.e., for any player $i\in \mathcal{I}$,
\begin{align*}
u_i(\textbf{s}=\textbf{s}^{\star,NE})
 \ge u_i([s_i=s,\textbf{s}_{-i}=\textbf{s}^{\star,NE}_{-i}]), \; \forall s \in \Delta(\mathcal{A}_i).
\end{align*}
The respective optimal utility value for each player is denoted by $u^{\star,NE}_i$.
\end{definition}

Regarding the non-cooperation among the overall players, the NE assumes that players choose actions independently, i.e., $\pr(\textbf{a}|\textbf{s}) = \prod_{i\in\mathcal{I}} s_i(a_i)$ in (\ref{eqn:expected utility}). However, it is possible to extend the sets of strategies available to the players by allowing them to correlate their choices. Motivated by that, a more general concept than NE, called the correlated equilibrium (CE), is proposed. In CE, the players can receive recommendations on what to play from an omniscient mediator. To be specific, at time $k$, the imagined mediator samples an $N$-tuple joint action $\textbf{a} = \{a_1,\ldots,a_N\}$ as a mode of play w.p. $\textbf{s}(\textbf{a})$, and the recommended action for player $i$ is $a_i=a$. Player $i$ may accept the recommendation or may use a meta-strategy, denoted by a transition $t(a_{i}): \mathcal{A}_i \rightarrow \mathcal{A}_i$, when it is suggested to play $a_{i}$. At a CE, no such meta-strategy would improve each player's expected utility if the others are assumed to play according to the recommendation. Hence, we have the following definition.
\begin{definition}[Correlated Equilibrium]\label{def:CE}
For this game $\mathcal{G}$, a strategy profile $\textbf{s}^{\star,CE}$ is a correlated equilibrium if and only if
\begin{equation}\label{eqn:correlated equilibrium}
\begin{aligned}
  &\sum_{\textbf{a}^{-} \in \mathcal{A}_{-i}} \pr(\textbf{a}_{-i}=\textbf{a}^{-}|a_{i}=a,\textbf{s}^{\star,CE}) \\
 &\{u_{i}( [a_{i}=a, \textbf{a}_{-i}=\textbf{a}^{-}])-u_{i}( [t(a), \textbf{a}_{-i}=\textbf{a}^{-}])\} \geq 0,
\end{aligned}
\end{equation}
for all players, all $a \in \mathcal{A}_i$ s.t. $\pr(a_i=a|\textbf{s}^{\star,CE})>0$, and all transitions $t(a) \in \mathcal{A}_i$.
We denote by $u^{\star,CE}_{i}$ the corresponding optimal utility value for each player.
\end{definition}

The CE concept can deal with some drawbacks of the original NE concept. One of the advantages of CE is that it can be computed in polynomial time (via a linear programming); whereas, the respective complexity for NE computation (finding its fixed point completely) is known as an NP-hard problem, see~\cite{nisan2007algorithmic}. More importantly, at a CE, multiple self-interested players may achieve higher rewards by coordinating their actions than they could at an NE. The comparison between NE and CE is analyzed rigorously in the following section.

\section{Main Results}\label{section:main result}

In this section, we discuss the equilibrium solution for this game under two different cases: with or without energy constrains. The specific representations and the comparison between NE and CE are also provided.

\subsection{Without Energy Constraints}

First, we consider the existence and uniqueness of NE for Prob. \ref{prob:game problem}, and provide the complete NE solution in the following theorem.

\begin{theorem}
The multi-player non-cooperative game $\mathcal{G}$ admits a unique NE, which has the property that all players transmit at its corresponding maximum power level; i.e.,
\begin{align*}
 s^{\star,NE}_i(a_i) =\left\{
          \begin{array}{ll}
            1, & \hbox{if $a_i=e^{(m_i)}_i$;} \\
            0, & \hbox{others.}
          \end{array}
        \right. \; \forall i\in \mathcal{I}.  \addtag\label{eqn:NE result}
\end{align*}
\end{theorem}

\begin{pf}
First, we consider the existence of a pure strategy NE, denoted by $\{a^{\star}_i,\textbf{a}^{\star}_{-i}\}$. By definition, we have
\begin{align*}
 &a^{\star}_i = \arg \max_{a_i \in \mathcal{A}_i} \; u_i([a_i,\textbf{a}^{\star}_{-i}]) \\
 \Leftrightarrow  &a^{\star}_i =\arg \min_{a_i} \; f(\gamma_i(a_i,\textbf{a}^{\star}_{-i})) = \arg \min_{a_i} \; f( \frac{L a_i}{\hat{q}^{\star}_{-i}+\sigma^2})\\
 \stackrel{(1)}{\Leftrightarrow} &a^{\star}_i = e^{(m_i)}_{i}.
\end{align*}
The derivation of $(1)$ is based on that $f(\cdot)$ decreases with $a_i$. Hence, we obtain a pure strategy NE shown in (\ref{eqn:NE result}).

Next, we prove the uniqueness of this NE. If there exists another NE, denoted by $\tilde{s}^{\star}_i$ with $\tilde{s}^{\star}_i(e^{(m_i)}_i)<1$,
\begin{align*}
u_i(\tilde{\textbf{s}}^{\star})&=
\sum_{\textbf{a}_{-i}\in\mathcal{A}_{-i}} \sum_{a_i\in\mathcal{A}} [\prod_{j\neq i} \tilde{s}^{\star}_j(a_j)] \tilde{s}^{\star}_i(a_i) u_i([a_i,\textbf{a}_{-i}]) \\
& < \sum_{\textbf{a}_{-i}\in\mathcal{A}_{-i}} \sum_{a_i\in\mathcal{A}}[\prod_{j\neq i} \tilde{s}^{\star}_j(a_j)] \tilde{s}^{\star}_i(a_i) u_i([e^{(m_i)}_i,\textbf{a}_{-i}]) \\
&= \sum_{\textbf{a}_{-i}\in\mathcal{A}_{-i}} [\prod_{j\neq i} \tilde{s}^{\star}_j(a_j)] u_i([e^{(m_i)}_i,\textbf{a}_{-i}]) = u_i([s^{\star}_{i},\tilde{\textbf{s}}^{\star}_{-i}]).
\end{align*}
This shows that player $i$ tends to adopt $s^{\star}_{i}$ if the others keep their strategy unchanged, which contradicts the NE concept. Hence, $s^{\star,NE}_i$ is unique.
$\hfill \blacksquare $
\end{pf}

Recall that, a CE is a joint distribution over actions from which no agent is motivated to deviate unilaterally. The following theorem demonstrates the uniqueness of the CE solution, and captures the relationship between CE and NE for the game $\mathcal{G}$.

\begin{theorem}
The multi-player game $\mathcal{G}$ has a unique CE, denoted by $\textbf{s}^{\star,CE}$, and it is equivalent to the NE.
\end{theorem}

\begin{pf}
From the definition of CE and conditional probability, we have
\begin{align*}\addtag \label{eqn:CE extention}
\sum_{\textbf{a}_{-i}} \textbf{s}^{\star,CE}(\textbf{a}) [u_{i}(\textbf{a})-u_{i}([t(a_i),\textbf{a}_{-i}])] \geq 0, \;\forall a_i, t(a_i)\in\mathcal{A}_i,
\end{align*}
in which $\textbf{s}^{\star,CE}(\textbf{a})$ is the probability over joint action set $\textbf{a}$ following the CE strategy profile, and $t(\cdot)$ is a mapping from $\mathcal{A}_{i}$ to $\mathcal{A}_{i}$. Next, we will interpret the computation of $\textbf{s}^{\star,CE}$ based on (\ref{eqn:CE extention}).

For player $1$, if $a_{1} = e^{(1)}_1$, then at a CE, for all $a_1, t(a_1)\in\mathcal{A}_1$ the following inequality holds:
$$\sum_{\textbf{a}_{-i}} \textbf{s}^{\star,CE}([e^{(1)}_1,\textbf{a}_{-1}]) [u_{1}([e^{(1)}_1,\textbf{a}_{-1}])-u_{i}([t(e^{(1)}_1),\textbf{a}_{-1}])] \geq 0. $$
Based on property (1) in the Appendix, we have $u_{1}([e^{(1)}_1,\textbf{a}_{-1}])< u_{1}([t(e^{(1)}_1),\textbf{a}_{-1}]),\; \forall \textbf{a}_{-1} \in \mathcal{A}_{-1}$ and $t(a_1)\neq a_1$. Therefore, $\textbf{s}^{\star,CE}([e^{(1)}_1,\textbf{a}_{-1}]) = 0, \forall \textbf{a}_{-1} \in \mathcal{A}_{-1}$.
Analogously, we can obtain that $\textbf{s}^{\star,CE}([e^{(1)}_i,\textbf{a}_{-i}]) = 0, \forall \textbf{a}_{-i} \in \mathcal{A}_{-i},\forall i \in \{1,\cdots,N\}$. That is, all players will choose their smallest power level $e^{(1)}_i$ w.p. $0$, no matter what strategies the others take.

Here, we construct a game $\mathcal{G}'$ with the action space excluding the lowest power levels, i.e., $\mathcal{A}'_{i} = \{e^{(2)}_i,\cdots,e^{(m_i)}_i\}, \forall i\in \mathcal{I}$. Obviously, the CE problem of the original game $\mathcal{G}$ is converted into that of the game $\mathcal{G}'$. Let $\textbf{s}'^{\star,CE}$ denote the CE of $\mathcal{G}'$.
Similar to the aforementioned analysis, we can obtain $\textbf{s}'^{\star,CE}([e^{(2)}_i, \textbf{a}_{-i}]) = 0, \forall \textbf{a}_{-i} \in \mathcal{A}'_{-i}$ and $\forall i\in\mathcal{I}$.
Consequently, the CE of the original game $\mathcal{G}$ can be computed easily by recursive analysis:
\begin{equation}\label{eqn:CE result}
  \textbf{s}^{\star,CE}(\textbf{a}) = \left\{\begin{array}{ll} 1,& \mathrm{if\;} \textbf{a}=\{e^{(m_1)}_1,\ldots,e^{(m_N)}_N\}.\\ 0,& \mathrm{others}.\end{array}\right.
\end{equation}
By comparing (\ref{eqn:NE result}) and (\ref{eqn:CE result}), the equivalence between the CE and NE is proved.
$\hfill \blacksquare $
\end{pf}

\subsection{With Energy Constraint}

As discussed previously, the optimal response for each player is to transmit the estimation packet constantly at its maximum power level. Nevertheless, such inefficient situation is less achievable, especially in a practical CPS, as the energy for sensor transmission is restricted. Here, we provide different constraints on the expected power consumption for each sensor, and provide the constrained game as follows:
\begin{problem}\label{prob:constained game}
For each player $i\in\mathcal{I}$,
\begin{align*}
\max_{\textbf{s}\in \Delta(\mathcal{A})} \; &u_{i}(\textbf{s}) \\
s.t. \; & \sum_{\textbf{a} \in\mathcal{A}} \textbf{s}(\textbf{a}) =1, \\
        & \E[a_i|s_i] = \sum_{a_i\in\mathcal{A}_i} s_i(a_i)a_i \leq e^{\max}_i. \addtag\label{eqn:energy constraint}
\end{align*}
Moreover, a strict power constraint is considered: $e^{(1)}_i<e^{\max}_i < e^{(m_i)}_i$.
\end{problem}

For the NE solution, we have the following result.

\begin{theorem}\label{them:NE result}
The constrained game (i.e., Prob. \ref{prob:constained game}) has a unique mixed strategy NE with its specific representation given in (\ref{eqn:NE result for constrained game}).
\end{theorem}

\begin{pf}
Let $\textbf{s}^{\star,NE} = \{s^{\star,NE}_{1},\ldots,s^{\star,NE}_{N}\}$ denote the optimal strategy profile.
For player $i$, given the optimal strategies of others $\textbf{s}^{\star,NE}_{-i}$, we have
\begin{align*}
  u_i(s^{\star,NE}_{i},\textbf{s}^{\star,NE}_{-i})=&\max_{s_{i} \in \Delta(\mathcal{A}_i)} \; u_i(s_{i},s^{\star,NE}_{-i}) \\
  =& \max_{s_{i}} \sum^{m_i}_{l=1} s_{i}(a_i=e^{(l)}_i) u^{(l)}_i, \addtag \label{eqn:transformed optimization}
\end{align*}
in which
\begin{align*}\addtag\label{eqn:added notation}
u^{(l)}_i
&\triangleq u_i(a_i=e^{(l)}_i,\textbf{s}^{\star}_{-i}) \\
& =
\sum_{\textbf{a}^{-}\in \mathcal{A}_{-i}} \prod^{N}_{j=1,j\neq i} s^{\star}_{j}(a_j) u_i(a_i = e^{(l)}_i,\textbf{a}^{-}) \\
&  < \sum_{\textbf{a}^{-}\in \mathcal{A}_{-i}} \prod^{N}_{j=1,j\neq i} s^{\star}_{j}(a_j) u_i(a_i = e^{(l')}_i,\textbf{a}^{-})\\
& = u^{(l')}_i,
\end{align*}
and in which $l' \in \{l+1, \ldots, m_i\}$.

Note that maximum utility is achieved when the equality in (\ref{eqn:energy constraint}) holds. Next, we rewrite the energy constraint in (\ref{eqn:energy constraint}) by replacing $s_i(e^{(1)}_i)$ with $1- \sum^{m_i}_{l=2} s_i(e^{(l)}_i)$. We can obtain that
\begin{align*}
  \sum^{m_i}_{l=2} d_l s_i(e^{(l)}_i) =1,
\end{align*}
in which $d_l \triangleq \frac{e^{(l)}_i-e^{(1)}_i}{e^{\max}_i-e^{(1)}_i}>0, \forall l \in \{2, \ldots, m_i\}$ is a constant,
and
\begin{align*}\addtag\label{eqn:maximization problem}
  &\max_{s_{i}} \; u_i(s_{i},s^{\star,NE}_{-i}) \\
  = &\max_{s_{i}} \; u^{(1)}_{i}+\sum^{m_i}_{l=2}\frac{u^{(l)}_i-u^{(1)}_i}{d_l}\cdot [d_l s_i(e^{(l)}_i)].
\end{align*}

According to property (2) in the Appendix, we have
\begin{align*}
  0<\frac{u^{(l)}_i-u^{(1)}_i}{d_l}< \frac{u^{(l')}_i-u^{(1)}_i}{d_{l'}}
\end{align*}
in which $l' \in \{l+1, \ldots, m_i\}$.
Apparently, the optimal solution is obtained when $d_{m_i} s_i(e^{(m_i)}_i) =1$, and the formulation of NE for this constrained game is: for all players,
\begin{align*}
 s_i^{\star,NE}(a_i) =
 \left\{\begin{array}{lll} (d_{m_i})^{-1}=\frac{e^{\max}_i-e^{(1)}_i}{e^{(m_i)}_i-e^{(1)}_i},& \mathrm{if\;} a_i= e^{(m_i)}_i,\\ 1-(d_{m_i})^{-1}=\frac{e^{(m_i)}_i-e^{\max}_i}{e^{(m_i)}_i-e^{(1)}_i},& \mathrm{if\;} a_i = e^{(1)}_i,\\ 0,& \mathrm{others}\addtag\label{eqn:NE result for constrained game}.\end{array}\right.
\end{align*}
The uniqueness of the NE is guaranteed by the optimal solution of (\ref{eqn:maximization problem}).
$\hfill \blacksquare $
\end{pf}

Note that at the NE, each player transmits the data packet with the highest or lowest energy power, regardless of the middle levels. Motivated by this, we propose the following mechanism.
\begin{definition}\label{def:correlated policy}
  We define the set of correlation policies as follows:
  \begin{itemize}
    \item Assume at each time $k$, all sensors can observe a signal in the form of a random variable $X(k)$, uniformly distributed over the integers $\{1,\ldots,N\}$.
    \item A correlated strategy of sensor $i$ is described by two numbers: $\alpha_i\in[0,1]$ and $\beta_i\in[0,1]$.
    \item At time $k$, if $X(k)=i$, then sensor $i$ is chosen to transmit a packet at the highest power $e^{(m_i)}_i$ w.p. $\alpha_i$ and at the lowest power $e^{(1)}_i$ w.p. $(1-\alpha_i)$. Otherwise it transmits w.p. $\beta_i$ for the highest power $e^{(m_i)}_i$ and w.p. $(1-\beta_i)$ for the lowest.
  \end{itemize}
\end{definition}

Next, we interpret the computation of the CE under this correlated policy. To simplify the calculation and obtain a closed-form formula of the CE, we assume that $d_{m_i}=d_{m_j},\forall i\neq j$ for this constrained game.
Recall that, the CE concept, compared with the ordinary NE, can simultaneously improve the benefit of each player under this competitive environment. The following theorem captures this and provides a complete representation of the CE.

\begin{theorem}\label{them:CE result}
This constrained game with $d_{m_i}=d,\forall i\in \mathcal{I}$, admits the existence of a CE under the correlation policy proposed in Def. \ref{def:correlated policy}. The corresponding parameter is illustrated by (\ref{eqn:CE paramter}).
Moreover, the CE outcome of this game is superior to the NE outcome for each player $i\in\mathcal{I}$.
\end{theorem}

\begin{pf}
First, we discuss the expected utility for player $i$ under this policy. By definition, if $X(k)=i$, then
$$s_i(a_i)= \mathds{1}_{\{e^{(m_i)}_i\}}(a_i)\alpha_i+\mathds{1}_{\{e^{(1)}_i\}}(a_i)(1-\alpha_i),$$ and for $j\neq i$
$$s_j(a_j)= \mathds{1}_{\{e^{(m_j)}_j\}}(a_j)\beta_j+\mathds{1}_{\{e^{(1)}_j\}}(a_j)(1-\beta_j).$$ The expected utility for the sensor $i$ is:
\begin{align*}
  u_i(\textbf{s}) &= \alpha_i u_i([e^{(m_i)}_i,s_l,\textbf{s}_{-i-l}]) + (1-\alpha_i) u_i([e^{(1)}_i,s_l,\textbf{s}_{-i-l}]),
\end{align*}
in which $l\neq j$, and the definition of $u_i([a_i,s_l,\textbf{s}_{-i-l}])$ is similar to (\ref{eqn:added notation}).

Analogously, if $X(k)=l$, the expected utility for sensor $i$ is:
\begin{align*}
  u_i(\textbf{s}') &= \beta_i u_i([e^{(m_i)}_i,s'_l,\textbf{s}_{-i-l}]) + (1-\beta_i) u_i([e^{(1)}_i,s'_l,\textbf{s}_{-i-l}]),
\end{align*}
in which
$$s'_l(a_l)= \mathds{1}_{\{e^{(m_l)}_l\}}(a_l)\alpha_l+\mathds{1}_{\{e^{(1)}_l\}}(a_l)(1-\alpha_l).$$

Next, we consider the computation of the CE. Solving the constrained optimization problem, as well as finding the CE, becomes rapidly intractable when the number of players $N$ increases. Here, we can restrict to the same strategy $(\alpha^{\star},\beta^{\star})$ being adopted by all players, and investigate if a single player deviating from this strategy by using a different strategy $(\alpha,\beta)$.

If $X(k)=i$ and player $i$ adopts a meta-strategy characterized by $\alpha$, then it obtains the benefit
\begin{align*}
 U_{\alpha}
  &=\sum_{\textbf{a}_{-i-l}} \pr(\textbf{a}_{-i-l}|\alpha^{\star},\beta^{\star}) \big[\alpha\beta^{\star}
  u^{m,m}_{i} +\alpha(1-\beta^{\star})u^{m,1}_i
 \\ &+ (1-\alpha)\beta^{\star}u^{1,m}_i
+ (1-\alpha)(1-\beta^{\star})
  u^{1,1}_{i}\big],
\end{align*}
in which, for example, $u^{m,m}_{i} \triangleq u_{i}([e^{(m_i)}_{i},e^{(m_l)}_{l},\textbf{a}_{-i-l}])$ and $u^{1,1}_{i} \triangleq u_{i}([e^{(1)}_{i},e^{(1)}_{l},\textbf{a}_{-i-l}])$.
%
%
%
If $X(k) = l$ and $\beta$ is used by player $i$, then its utility is
\begin{align*}
 U_{\beta} = & \sum_{\textbf{a}_{-i-l}} \pr(\textbf{a}_{-i-l}|\alpha^{\star},\beta^{\star})\big[\beta\alpha^{\star}
  u^{m,m}_{i}+ \beta(1-\alpha^{\star})u^{m,1}_i \\&+(1-\beta)\alpha^{\star}u^{1,m}_i
  + (1-\beta)(1-\alpha^{\star})u^{1,1}_{i}\big].
\end{align*}
Then, the expected utility of player $i$ by adopting $(\alpha,\beta)$ is
\begin{align*}
  u_i(\alpha,\beta)=& \frac{ U_{\alpha}+(N-1)U_{\beta}}{N} \\
 \stackrel{(1)}{=}& \sum_{\textbf{a}_{-i-l}} \pr(\textbf{a}_{-i-l}|\alpha^{\star},\beta^{\star})
   \frac{N-1}{N}(\alpha^{\star}-\beta^{\star})\hat{u}\beta +c',
\end{align*}
in which $\hat{u}= u^{m,m}_i-u^{m,1}_i-u^{1,m}_i+u^{1,1}_i<0$ is based on property (3) in the Appendix, and $c'$ is a constant. The equality $(1)$ is derived by replacing $\alpha$ with the expression $N d^{-1}-(N-1)\beta$ (obtained by the energy constraint). Hence, $u_i(\alpha,\beta)$ is an affine function in $\beta$, and the optimal $\beta$ is
\begin{align*}\addtag\label{eqn:CE paramter}
  \beta^{\star} =
\left\{
            \begin{array}{ll}
              \max(0, \frac{Nd^{-1}-1}{N-1}), & \hbox{If $\alpha^{\star}>\beta^{\star}$;}  \\
              \min(1, \frac{Nd^{-1}}{N-1}), & \hbox{If $\alpha^{\star}<\beta^{\star}$.}
            \end{array}
          \right.
\end{align*}
Last, we compare the CE outcome with that of the NE. The NE is a special case of the CE, that is $\alpha^{\star} = \beta^{\star} =s$. From the previous discussion, we know that
$\arg \max_{\beta} u_i(\alpha,\beta) = \beta^{\star} \neq s$. Hence, we have the payoff of the CE strictly greater than the payoff of the NE.
$\hfill \blacksquare $
\end{pf}

\begin{rem}
In the absent of energy limitations, the best response for each sensor is transmitting with the maximum energy level, no matter what equilibrium it chooses. {The result is of common sense since the players' utilities do not involve the power expenditure items. However, when taking the power restrictions into account, the correlation mechanism displays its advantages over the NE. Intuitively, in NE every player will take full advantage of their transmission power, which inevitably causes heavy communication conflicts. But, the existence of mediator in CE can coordinate the players' behaviors to alliterate these conflicts and achieve better performance.}
\end{rem}

%

\section{Simulation and Examples}\label{section:simulation and example}

In this section, we will compare the NE outcome and the CE outcome of the game-theoretic model using an example. Here, we consider a multi-agent system with three sensors and one remote estimator. The monitored dynamic processes have respective system parameters, demonstrated in Tab. \ref{tab:system parameters}. In addition, the error covariances of the Gaussian noises are $Q_i = R_i =0.8, \forall i \in \{1,2,3\}$. Some parameters of the communication channel are given in Tab. \ref{tab:system parameters}. Moreover, $e^{(1)}_i=0$ (note that each sensor may choose to stay inactive), $h_i =1,\forall i \in \{1,2,3\}$ and $L = 2$. For the SER, we adopt the formula: $f(\gamma) = 1- 2 \mathtt{Q}(\sqrt{4\gamma^{-1}-1})$, where $\mathtt{Q}(x) \triangleq \frac{1}{\sqrt{2\pi}}\int^{\infty}_{x} \exp(-\frac{y^2}{2})\d y$.

\begin{table}[]
\centering
\caption{System Parameters}
\label{tab:system parameters}
\begin{tabular}{l|l|l|l|l|l|l|l|l}
\hline
 & \multicolumn{4}{l|}{Process parameters} & \multicolumn{4}{l}{Channel parameters} \\ \hline
 Process 1& $A_1$ & 0.9  & $C_1$ & $1$   & $e^{(m_1)}_1$  & $1$  & $e^{\max}_1$  & $0.5$    \\ \hline
 Process 2&  $A_2$ & 0.8  & $C_2$ & $1.1$  & $e^{(m_2)}_2$ & $0.8$ &  $e^{\max}_2$ & $0.4$  \\ \hline
 Process 3&$A_3$ & 0.7  & $C_3$ & $1.2$   & $e^{(m_3)}_3$  & $0.6$  & $e^{\max}_3$   &   $0.3$  \\ \hline
\end{tabular}
\end{table}

By Theorem. \ref{them:NE result} and Theorem. \ref{them:CE result}, we can obtain the following two strategy profiles for the sensor to transmit data packets:
\begin{itemize}
  \item $\textbf{s}^{NE}: s_i(0)=s_i(e^{(m_i)}_i)=0.5, \; \forall i \in\{1,2,3\}$.
  \item $\textbf{s}^{CE}:        s_i(e^{(m_i)}_i)
      = \left\{
           \begin{array}{ll}                                                          0.75, & \hbox{if sensor $i$ is chosen;} \\
   0.25, & \hbox{others.}                                                 \end{array}                             \right.$
and $s_i(0)= 1-s_i(e^{(m_i)}_i)$.
\end{itemize}

\begin {figure}
\centering
\includegraphics[width=8.4cm]{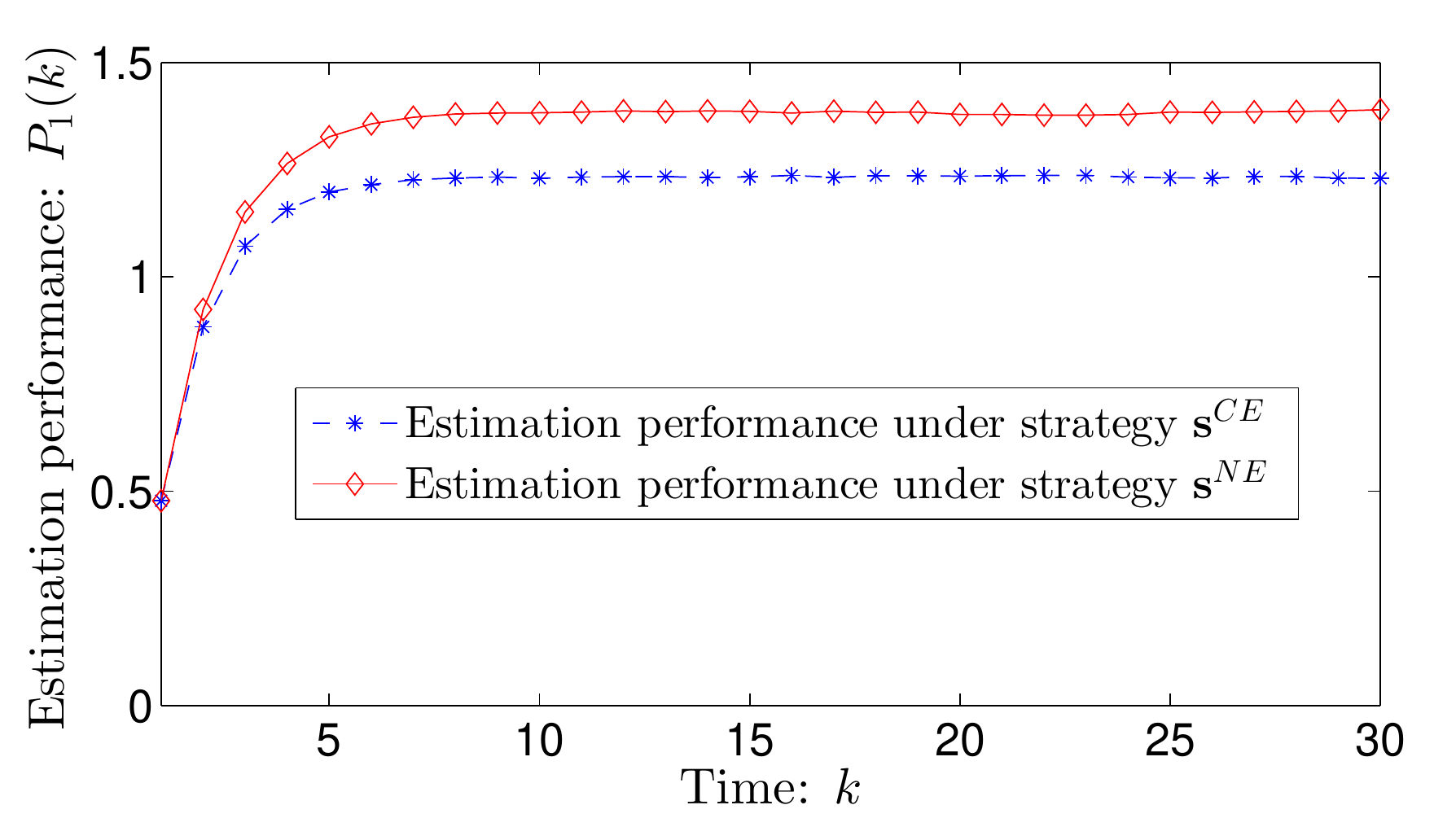}
\includegraphics[width=8.4cm]{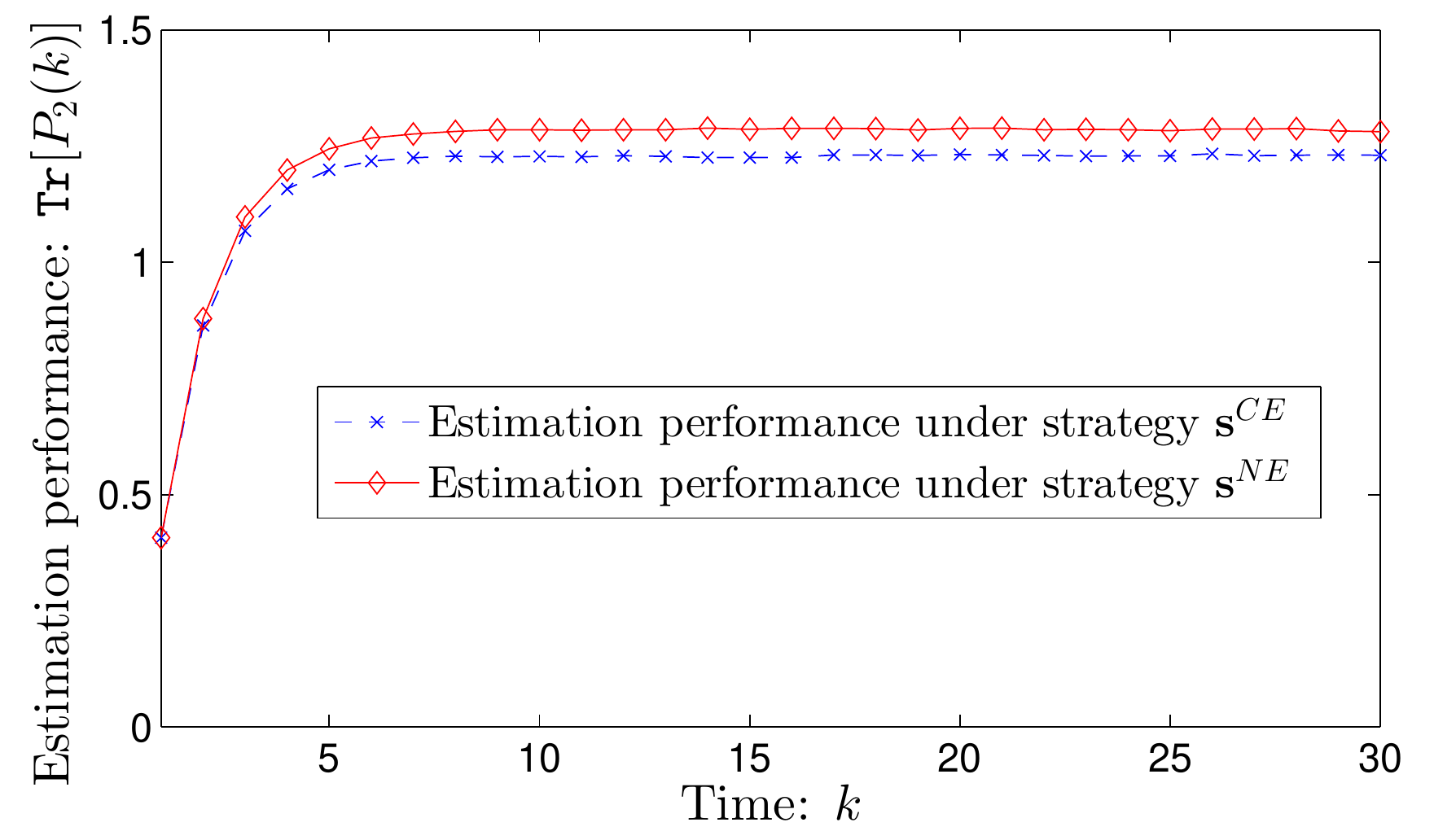}
\caption{Comparison between $\textbf{s}^{NE}$ and $\textbf{s}^{CE}$ for sensors 1 and 2.}\label{fig}
\end{figure}

\begin{figure}
\begin{center}
\includegraphics[width=8.4cm]{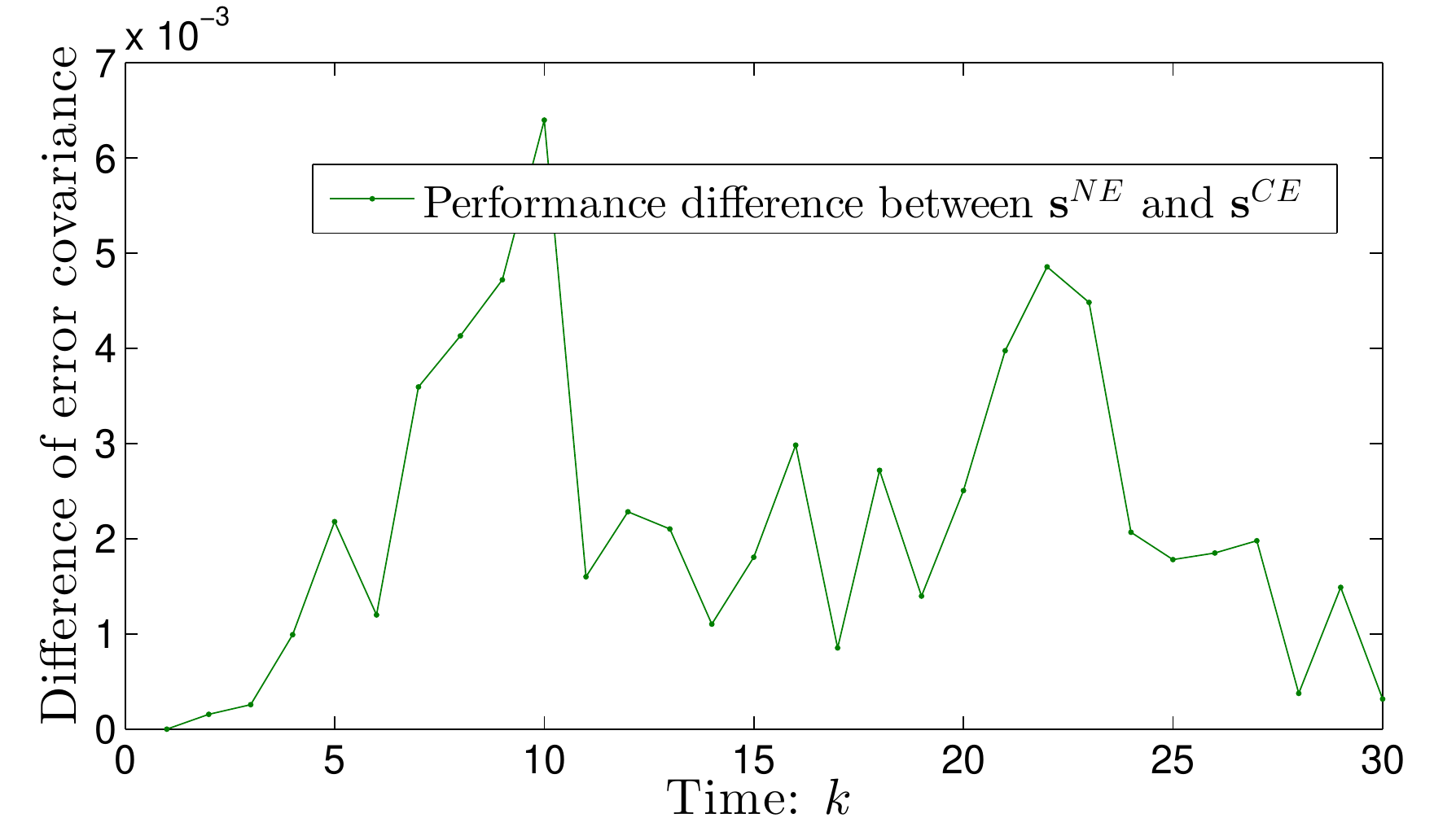}    
\caption{Performance difference between $\textbf{s}^{NE}$ and $\textbf{s}^{CE}$ for sensor $3$.}
\label{fig:3}
\end{center}
\end{figure}

Via $100,000$ Monte Carlo simulations, the comparison (between $\textbf{s}^{NE}$ and $\textbf{s}^{CE}$) results of state estimation error covariance for sensors 1 and 2 are depicted in Fig. \ref{fig}. Furthermore, we represent the performance difference of sensor 3 in Fig. \ref{fig:3}. All comparison results illustrate the analytical performance in Theorem. \ref{them:CE result} and highlight the superiority of the proposed coordination mechanism. Last, but not least, comparing Fig. \ref{fig} and Fig. \ref{fig:3}, the performance difference between the NE and the CE decreases as the energy constraint becomes stronger.


%
%

\section{Conclusion}\label{section:conclusion}

We have investigated the remote estimation issue for a multi-sensor system under the game-theoretic framework. Motivated by the concept of Nash equilibrium in the previous work, we analyzed the performance advantage brought by the correlation policy. In the absence of power constraints, the correlated equilibrium outcome is equal to the NE outcome. However, the correlated policy improves the estimation performance in the presence of power constraints.

\bibliography{ifacconf}             








\appendix
\section{Properties of the utility function}

Consider the utility function $u_i(a_i,\textbf{a}_{-i}), \forall i \in \set$, defined in (\ref{eqn:utility function}). We can obtain the following properties:
\begin{enumerate}
  \item $\frac{\partial u_i(a_i,\textbf{a}_{-i})}{\partial a_i}>0$ if $a_i>0$.
  \item If $a_2>a_1>a_0>0$, then
$$
w_{a_0}(a_i=a_2)>w_{a_0}(a_i=a_1),
$$
in which $w_{a_0}(a_i) \triangleq \frac{u_i(a_i,\textbf{a}_{-i})-u_i(a_i=a_0,\textbf{a}_{-i})}{a_i-a_0}
$ and $\textbf{a}_{-i}$ is given.
  \item If $a_2>a_1>0$ and $a_4>a_3>0$ then
\begin{align*}
u_i(a_2,a_4,\textbf{a}_{-i-l})&-u_i(a_2,a_3,\cdot)\\&-u_i(a_1,a_4,\cdot)
+u_i(a_1,a_3,\cdot)<0, \end{align*}
where $\textbf{a}_{-i-l}$ is given.
\end{enumerate}

\begin{pf}
\begin{enumerate}
  \item The first statement can be obtained from $c_i<0$ and
    $$
    \frac{\partial u_i(a_i,\textbf{a}_{-i})}{\partial a_i} = c_i f'(\gamma_i)\frac{\gamma_i}{a_i}.
     $$
  \item Since $f''(\gamma_i)< 0$, then
      $$ \frac{\partial^2 u_i(a_i,\textbf{a}_{-i})}{\partial a^2_i} = c_i f''(\gamma_i)(\frac{\gamma_i}{a_i})^{2}>0. $$
Hence, $u_i(a_i,\textbf{a}_{-i})$ is an increasing and strictly convex function in $a_i$. Obviously, we can obtain that if $a_2>a_1>a_0>0$, then
$$
w_{a_0}(a_2)>w_{a_0}(a_1).
$$
  \item Consider the second partial derivative of function $u_i(\cdot)$, $$
   \frac{\partial^2 u_i(a_i,a_l\textbf{a}_{-i-l})}{\partial a_i \partial a_l} = \frac{-c_i h_l\gamma^2_i}{h_i L a^2_i} [f'(\gamma_i)+\gamma_i f''(\gamma_i)] <0.
$$
Hence, we have
$$
\frac{u_i(a_2,a_4,\cdot)-u_i(a_1,a_4)}{a_2-a_1} < \frac{u_i(a_2,a_3,\cdot)-u_i(a_1,a_3)}{a_2-a_1}
$$
Obviously, the statement (3) holds.$\hfill \blacksquare $
\end{enumerate}

\end{pf}

\end{document}